\documentclass[twocolumn,aps,prd,groupedaddress,nofootinbib,showpacs]{revtex4}
\usepackage{bbm}
\usepackage[T1]{fontenc} 
\usepackage{graphicx}
\usepackage{amsmath}
\usepackage{bm}
\usepackage{slashed}
\usepackage{epstopdf}
\usepackage{epsfig}
\usepackage{verbatim}
\usepackage{mathrsfs}
\usepackage{bm}


\def\g0{\gamma_0}

\def\nn{\nonumber}
\def\beqa{\begin{eqnarray}}
\def\eeqa{\end{eqnarray}}
\def\beqn{\begin{eqnarray}}
\def\eeqn{\end{eqnarray}}
\def\beq{\begin{equation}}
\def\eeq{\end{equation}}

\def\e{\epsilon}

\def\I4{I_4}

\begin{document}

\preprint{DESY~18-153\hspace{12cm}ISSN~0418-9833}

\boldmath
\title{
{\rm DESY 18--153\hspace{12.5cm}ISSN 0418-9833}\\
{\rm September 2018\hspace{14.7cm}\phantom{x}}\\
Inclusive $\chi_{cJ}$ production in $\Upsilon$ decay at $\mathcal{O}(\alpha_s^5)$ in NRQCD factorization}
\unboldmath

\author{Zhi-Guo He, Bernd A. Kniehl, Xiang-Peng Wang}

\affiliation{{II.} Institut f\"ur Theoretische Physik, Universit\"at Hamburg,
Luruper Chaussee 149, 22761 Hamburg, Germany}

\date{\today}

\begin{abstract}
Inclusive $\chi_{cJ}$ $(J=0,1,2)$ production from $\Upsilon(1S)$ decay is
studied within the framework of nonrelativistic QCD (NRQCD) factorization at
leading order in $v_Q^2$, which includes the contributions of
$b\bar{b}({}^3S_1^{[1]})\to c\bar{c}(^3P_J^{[1]})+X$ and
$b\bar{b}({}^3S_1^{[1]})\to c\bar{c}(^3S_1^{[8]})+X$.
For both channels, the short-distance coefficients are calculated through
${\cal O}(\alpha_s^5)$, which is next-to-leading order for the second one.
By fitting to the measured $\Upsilon(1S)$ branching fractions to
$\chi_{c1}$ and $\chi_{c2}$, we obtain the color-octet long-distance matrix
element (LDME)
$\langle\mathcal{O}^{\chi_{c0}}({}^3S_1^{[8]})\rangle
=(4.04\pm0.47_{-0.34}^{+0.67})\times10^{-3}$~GeV$^3$, where the first error is
experimental and the second one due to the renormalization scale dependence, if
we use as input
$\langle\mathcal{O}^{\chi_{c0}}({}^3P_0^{[1]})\rangle=0.107$~GeV$^5$ as
obtained via potential-model analysis.
Previous LDME sets, extracted from data of prompt $\chi_{cJ}$ hadroproduction,
yield theoretical predictions that systematically undershoot or mildly
overshoot the experimental values of $\mathcal{B}(\Upsilon\to \chi_{cJ}+X)$.
\end{abstract}


\maketitle

\section{Introduction}
Heavy-quarkonium production serves as an ideal laboratory to study both the
perturbative and nonperturbative aspects of QCD due to the hierarchy of energy
scales $m_Q v^2\ll m_Qv\ll m_Q$, where $m_Q$ is the mass of the heavy quark $Q$
and $v$ is its relative velocity in the rest frame of the heavy meson.
The effective quantum field theory of nonrelativistic QCD (NRQCD)
\cite{Caswell:1985ui} endowed with the factorization conjecture of
Ref.~\cite{Bodwin:1994jh} is the default theoretical approach to study
quarkonium production and  decay.
This conjecture states that the theoretical predictions can be separated into
process-dependent short-distance coefficients (SDCs) calculated perturbatively
as expansions in the strong-coupling constant $\alpha_s$ and supposedly
universal long-distance matrix  elements (LDMEs), scaling with definite power
of $v$ \cite{Lepage:1992tx}.
In this way, the theoretical calculations are organized as double expansion in
$\alpha_s $ and $v$.

During the past two decades, the NRQCD factorization approach has celebrated
numerous remarkable successes in describing both the production and decay of
heavy quarkonium (see Refs.~\cite{Brambilla:2010cs,Brambilla:2014jmp} and
references therein for a review).
However, there are still some challenges in understanding charmonium
production, in particular for the $J/\psi$ meson.
Prompt $J/\psi$ production has been studied in various environments on both
experimental and theoretical sides.
To date, the SDCs are available at next-to-leading order (NLO) in $\alpha_s$
for the yield \cite{Zhang:2009ym,Ma:2008gq} and polarization \cite{Gong:2009kp} in $e^{+}e^{-}$ annihilation, the yield in two-photon collisions
\cite{Klasen:2004tz,Butenschoen:2011yh}, the yield \cite{Butenschoen:2009zy}
and polarization \cite{Butenschoen:2011ks} in photoproduction, the yield
\cite{Ma:2010yw,Butenschoen:2010rq} and polarization
\cite{Butenschoen:2012px,Chao:2012iv,Gong:2012ug,Shao:2014yta} in
hadroproduction, etc.
Different sets of LDMEs were obtained by fitting experimental data adopting
different strategies.
Unfortunately, none of them can explain all the experimental measurements,
which challenges the universality of the NRQCD LDMEs.
Recently, it has been found that all the LDME sets determined from $J/\psi$
production data result in NLO predictions that overshoot the $\eta_c$
hadroproduction data \cite{Butenschoen:2014dra}.   
 
Because the P-wave states $\chi_{cJ}$ $(\mathrm{J=1,2})$ have substantial
branching fractions to $J/\psi$ though $\chi_{cJ}\to J/\psi+\gamma$, their
production can also be measured, which provides an additional playground to
test the NRQCD factorization hypothesis.
Moreover, the feed-down contributions from the $\chi_{c1}$ and $\chi_{c2}$
mesons to the yield and polarization of prompt $J/\psi$ hadropoduction is
sizable \cite{Gong:2012ug,Ma:2010vd}.
Unlike for the $J/\psi$ meson, inclusive $\chi_{cJ}$ production has only been
studied in a few processes, for example hadroproduction
\cite{Gong:2012ug,Ma:2010vd,Li:2011yc,Shao:2014fca,Jia:2014jfa}, $e^{+}e^{-}$ 
annihilation \cite{Chen:2014ahh}, top-quark decay \cite{Chao:2016chm},
$B$ hadron decay \cite{Beneke:1998ks}, and $\eta_b$ meson decay
\cite{He:2009sm,Li:2010ze}. 

In this work, we will study another interesting process, namely inclusive
$\chi_{cJ}$ production through $\Upsilon(1S)$ decay. 
On the experimental side, thanks to the large number of $\Upsilon(1S)$ decay
events collected with the Belle detector, $102\times 10^6$, the value
$\mathcal{B}(\Upsilon(1S)\rightarrow \chi_{c1} + X)=(1.90\pm0.35)\times 10^{-4}$
has recently been obtained \cite{Jia:2016cgl}, which is more precise than the
previous result $(2.3\pm 0.7 )\times 10^{-4}$ extracted by analyzing
$21.2\times 10^6$ $\Upsilon(1S)$ decay events collected with the CLEO~III
detector~\cite{Briere:2004ug}.
As for $\mathcal{B}(\Upsilon(1S)\rightarrow \chi_{c2} + X)$, the combined
result $(2.8\pm 0.8 )\times 10^{-4}$ was reported by the Particle Data Group in
2016 \cite{Patrignani:2016xqp}.

This process was considered theoretically more than two decades ago in
Refs.~\cite{Trottier:1993ze,Cheung:1996mh}, where the contributions of
$c\bar{c}$ pairs in color singlet (CS) $({}^3P_J^{[1]})$ and color octet 
(CO) $({}^3S_1^{[8]})$ Fock states were computed at leading order (LO) in
$\alpha_s$.
Moreover, in Ref.~\cite{Trottier:1993ze}, a cut on the soft-gluon energy 
was introduced to regularized the infrared (IR) divergences. 

Many of the theoretical works mentioned above have shown that the effects of
NLO QCD corrections may be large.
Therefore, in this work, we calculate the NLO QCD corrections to inclusive
$\chi_{cJ}$ production by $\Upsilon$ decay within the framework of NRQCD 
factorization.
Note that we do not take into account the contributions from $b\bar{b}$ pairs
in CO states, since they are suppressed by $v_b^4$.
The remainder of this paper is organized as follows.
In Sec.~II, we describe how to calculate the relevant SDCs in detail.
In Sec.~III, we present the numerical results and compare them with the
available experimental measurements.
In Sec.~IV, we summarize our results.
In Appendix~A, we list the master integrals arising in the virtual corrections
through ${\cal O}(\epsilon)$ in the expansion parameter of dimensional
regularization.
In Appendix~B, we list the soft integrals arising in the real corrections
implemented with phase space slicing.

\section{Calculation of SDCs}
\subsection{NRQCD Factorization and Notations}
%

In the NRQCD factorization formalism, at LO in $v_b$ and $v_c$, the decay width
of $\Upsilon\rightarrow \chi_{cJ} + X $ can be written as
\beqn
\lefteqn{\Gamma (\Upsilon \rightarrow \chi_{cJ} + X)
 =\langle \Upsilon | \mathcal{O}(^3S_1^{[1]}) |\Upsilon\rangle}\nn\\
&&{}\times \left[
  \hat{\Gamma}_1 \left(b\bar{b}({}^3S_1^{[1]})\rightarrow c\bar{c}({}^3P_J^{[1]}) + X\right)\langle\mathcal{O}^{\chi_{cJ}}({}^3P_J^{[1]})\rangle\right.\nn\\
&&{}+\left.  
  \hat{\Gamma}_8 \left(b\bar{b}({}^3S_1^{[1]})\rightarrow c\bar{c}({}^3S_1^{[8]}) + X\right)\langle\mathcal{O}^{\chi_{cJ}}(^3S_1^{[8]})\rangle
  \right],\qquad
\eeqn
where $\hat{\Gamma}_1$ and $\hat{\Gamma}_8$ are the SDCs and
$\langle \Upsilon | \mathcal{O}({}^3S_1^{[1]}) |\Upsilon\rangle$,
$\langle \mathcal{O}^{\chi_{cJ}}({}^3P_J^{[1]})\rangle$, and
$\langle \mathcal{O}^{\chi_{cJ}}({}^3S_1^{[8]})\rangle$ are the LDMEs.
We adopt the conventions for the LDMEs introduced in Ref.~\cite{Bodwin:1994jh}.
The LDMEs for $\chi_{cJ}$ production satisfy the multiplicity relations
\beqn
\langle\mathcal{O}^{\chi_{cJ}}({}^3P_J^{[1]})\rangle &=&
(2J+1)\langle\mathcal{O}^{\chi_{c0}}(^3P_0^{[1]})\rangle,
\nonumber\\
\langle\mathcal{O}^{\chi_{cJ}}({}^3S_1^{[8]})\rangle &=&
(2J+1)\langle\mathcal{O}^{\chi_{c0}}({}^3S_1^{[8]})\rangle,
\eeqn
which follow from heavy-quark spin symmetry at LO in $v_c$.

At LO in $\alpha_s$, at $\mathcal{O}(\alpha_s^4)$, only the CO subprocess
$b\bar{b}({}^3S_1^{[1]})\rightarrow c\bar{c}({}^3S_1^{[8]})+gg$ contributes, 
and its NLO QCD corrections include both virtual and real corrections at
$\mathcal{O}(\alpha_s^5)$.
CS contributions start to contribute at $\mathcal{O}(\alpha_s^5)$ via
$b\bar{b}({}^3S_1^{[1]})\to c\bar{c}({}^3P_J^{[1]})+ggg$.
We will calculate this consistently in dimensional regularization, with
$D=4-2\epsilon$ space-time dimensions.

Moreover, inclusive $J/\psi$ production by $e^{+}e^{-}$ annihilation
\cite{Abe:2002rb,Liu:2003jj,Zhang:2006ay,Gong:2009ng}  and $\Upsilon$
decay \cite{He:2009by} is known to receive substantial contributions
from events that also contain an open $c\bar{c}$ pair in the final state.
This motivates us to include the contributions from
$b\bar{b}({}^3S_1^{[1]})\rightarrow c\bar{c}({}^3P_J^{[1]}) + c\bar{c}g$ and
$b\bar{b}({}^3S_1^{[1]})\rightarrow c\bar{c}({}^3S_1^{[8]}) + c\bar{c}g$ at
$\mathcal{O}(\alpha_s^5)$ as well.
A similar study of $J/\psi$ production by $\Upsilon$ decay in the NRQCD
factorization approach will be presented in a forthcoming paper \cite{Jpsipaper}
and compared with previous investigations
\cite{Trottier:1993ze,Cheung:1996mh,He:2009by,Napsuciale:1997bz,He:2010cb}.

\begin{figure}
\begin{center}
\includegraphics[width=0.9\linewidth]{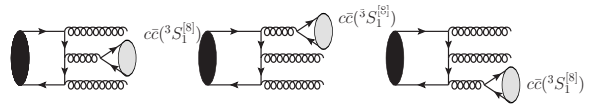}
\caption{Typical tree-level Feynman diagrams for the partonic subprocess
  $b\bar{b}({}^3S_1^{[1]})\rightarrow c\bar{c}({}^3S_1^{[8]}) + gg$.}
\label{ggtree}
\end{center}
\end{figure}

\begin{figure}
\begin{center}
\includegraphics[width=0.9\linewidth]{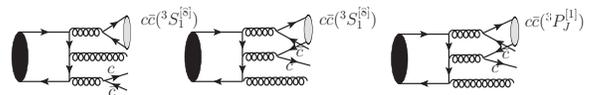}
\caption{Typical tree-level Feynman diagrams for the partonic subprocess
$b\bar{b}({}^3S_1^{[1]})\rightarrow c\bar{c}({}^3S_1^{[8]}/^3P_J^{[1]})+c\bar{c}g$.}
\label{ccgtree}
\end{center}
\end{figure}

There are 6 Feynman diagrams for the partonic subprocess
$b\bar{b}({}^3S_1^{[1]})\rightarrow c\bar{c}({}^3S_1^{[8]}) + gg$ at tree level
(see Fig.~\ref{ggtree}). 
The analytical and numerical results were first presented in
Ref.~\cite{Cheung:1996mh}, and we reproduce them.
To compute the NLO QCD corrections, the analytical expression for the scattering
amplitude must also be obtained in $D$ dimensions.
Ultraviolet (UV) and IR divergences are encountered in the
calculation of the NLO QCD corrections to
$b\bar{b}({}^3S_1^{[1]})\rightarrow c\bar{c}({}^3S_1^{[8]}) + gg$.
IR divergences also appear in the phase space integration of the subprocesses
$b\bar{b}({}^3S_1^{[1]})\rightarrow c\bar{c}({}^3P_J^{[1]}) + ggg$.
We will present in detail the treatment of these divergences in dimensional
regularization in the next three subsections.
The calculation of the associated open-charm subprocesses (Fig.~\ref{ccgtree})
is straightforward because no divergences appear in the phase space
integrations.
Their analytical expressions are too lengthy to be displayed here, and we thus
only present the numerical results for them in the next section.

In our analytical computation, the Feynman diagrams are generated using
FeynArts \cite{Hahn:2000kx}.
Algebraic operations such as color and Dirac algebra are performed with
FeynCalc \cite{Mertig:1990an} and FORM \cite{Kuipers:2012rf}.
For the virtual corrections, we use the Mathematica package \$Apart
\cite{Feng:2012iq} to decompose linearly dependent propagators in the loop
integrals to irreducible ones.
The latter are then reduced to master scalar integrals using the FIRE
\cite{Smirnov:2008iw} package.
The master integrals are evaluated numerically using the C$++$ package QCDLOOP
\cite{Carrazza:2016gav}.
Finally, the phase space integrations are performed numerically with the help
of the CUBA \cite{Hahn:2004fe} library. 

\subsection{Virtual Corrections}

\begin{figure}
\begin{center}
\includegraphics[width=0.9\linewidth]{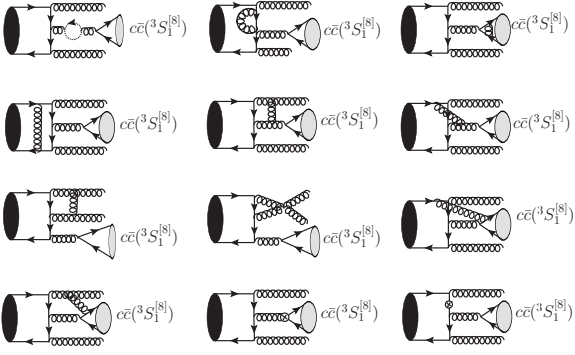}
\caption{Typical Feynman diagrams for the virtual corrections to the partonic
  subprocess  $b\bar{b}({}^3S_1^{[1]})\rightarrow c\bar{c}({}^3S_1^{[8]}) + gg$.}
\label{VC}
\end{center}
\end{figure}

Typical Feynman diagrams of the virtual corrections are shown in Fig.~\ref{VC}.
They fall into four groups, including self-energy diagrams, 
vertex correction diagrams, counterterm diagrams, and diagrams that are
generated from tree-level diagrams of Fig.~\ref{ggtree} by attaching 
one virtual-gluon line in all other possible ways. 

In the analytical calculation of the tree-level subprocess
$b\bar{b}$ $({}^3S_1^{[1]})(P)\rightarrow c\bar{c}({}^3S_1^{[8]})(K) +
g(k_3)g(k_4)$ 
and the corresponding virtual corrections, the Mandelstam variables are defined
as
\beqn
s \equiv  (P -k_4)^2,\  t \equiv  (P - K)^2, u  \equiv & (P - k_3)^2,
\eeqn
so that $s+u+t=4m_b^2+4m_c^2$, where $m_c$ and $m_b$ are the masses of the charm
and bottom quarks, respectively.
For convenience, we also introduce
\beqn
&& s_c\equiv s-4m_c^2,\qquad s_b\equiv 4m_b^2-s,\nn\\
&& u_c\equiv u-4m_c^2,\qquad u_b\equiv 4m_b^2-s.
\eeqn

Labeling the tree-level amplitude and the amplitude of the virtual corrections
as $\mathcal{M}_{\text{Born}}$ and $\mathcal{M}_{\text{virtual}}$, respectively, the
SDC of the virtual corrections is evaluated as
\beqn
d\hat{\Gamma}^{\text{VC}}= \frac{1}{4m_b}d \text{PS}_{1\rightarrow 3}\overline{\sum}
2\text{Re}(\mathcal{M}_{\text{Born}}^*\mathcal{M}_{\text{virtual}}), 
\eeqn
where $d \text{PS}_{1\rightarrow 3}$ is the three-body phase space and
$\overline{\sum}$ implies average over the color and polarization states.

The UV divergences are removed through renormalization.
We adopt a mixed renormalization scheme \cite{Klasen:2004tz}, in which the
renormalization constants $Z_2$, $Z_m$, and $Z_3$ of the heavy-quark field
$\psi_Q$, heavy-quark mass $m_Q$, and gluon field $A_{\mu}^{a}$ are defined in
the on-shell (OS) scheme, while the renormalization constant $Z_g$ of the
strong-coupling $g_s$ is defined in modified minimal-subtraction 
($\overline{\text{MS}}$) scheme.
At the one-loop level, they read
\beqn
\delta Z_g^{\overline{\text{MS}}} &=& -\frac{\beta_0}{2}\frac{\alpha_s}{4\pi}C_\epsilon\frac{1}{\epsilon_{\text{UV}}},\\
\delta Z_2^{\text{OS}} &=& -C_F\frac{\alpha_s}{4\pi}C_\epsilon\left[\frac{1}{\epsilon_{\text{UV}}}+\frac{2}{\epsilon_{\text{IR}}}+3\ln{\frac{\mu^2}{m^2}}+4
\right],
\label{field}\\
\delta Z_3^{\text{OS}} &=& \frac{\alpha_s}{4\pi}C_\epsilon(\beta_0-2C_A)\left[\frac{1}{\epsilon_{\text{UV}}}-\frac{1}{\epsilon_{\text{IR}}}\right],\\
\delta Z_m^{\text{OS}} &=& -3C_F\frac{\alpha_s}{4\pi}C_\epsilon\left[\frac{1}{\epsilon_{\text{UV}}}+\ln{\frac{\mu^2}{m^2}}+\frac{4}{3}\right]\label{mass},
\eeqn
where $C_\epsilon = (4\pi e^{-\gamma_E})^\epsilon$, $\mu$ is the renormalization
scale, and $\beta_0=(11/3)C_A -(4/3)T_F n_f$ is the one-loop coefficient of the
QCD beta function.
We have $n_f=3$ active quark flavors in our calculation.
In Eqs.~(\ref{field}) and (\ref{mass}), $m$ is to be substituted by $m_c$ and
$m_b$ for charm and bottom, respectively.

At the end of the reduction done with FIRE \cite{Smirnov:2008iw}, the loop
corrections are expressed in terms of some master integrals.
Among them, the coefficients of some tadpole and bubble integrals carry extra
poles $\frac{1}{\epsilon}$.
Therefore, we must calculate these integrals through $\mathcal{O}(\epsilon)$.
The results are given in Appendix~A.
We extract the divergences analytically and find that they cancel after
combination with the real corrections, which are IR divergent.

\subsection{Real Corrections}
\begin{figure}
\begin{center}
\includegraphics[width=0.9\linewidth]{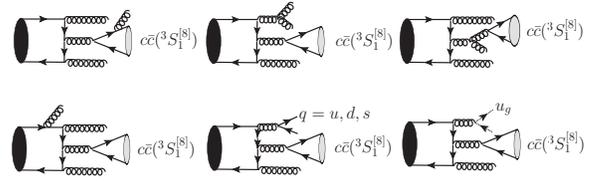}
\caption{Typical Feynman diagrams for the real corrections to the partonic
subprocess $b\bar{b}({}^3S_1^{[1]})\rightarrow c\bar{c}({}^3S_1^{[8]}) + gg$.}
\label{RC}
\end{center}
\end{figure}

In our calculation of the real corrections, we choose Feynman gauge with the
polarization sum of the gluon given by
\beqa
\sum_{\text{pol}^*}\epsilon_\mu\epsilon_\nu^*=-g_{\mu\nu}.
\eeqa
Consequently, the nonphysical degrees of freedom due to the gluon-ghost
contributions should be subtracted as
 \beqn
\lefteqn{\sum_{\text{col, pol}} |\mathcal{M}(b\bar{b}({}^3S_1^{[1]})\rightarrow c\bar{c}({}^3S_1^{[8]}) + ggg)|^2}\nn\\
&=&\sum_{\text{col, pol}^*} \left(|\mathcal{M}(b\bar{b}({}^3S_1^{[1]})\rightarrow c\bar{c}({}^3S_1^{[8]}) + ggg)|^2\right.\nn\\
&&{}-2 |\mathcal{M}(b\bar{b}({}^3S_1^{[1]})\rightarrow c\bar{c}({}^3S_1^{[8]}) + gu_g(k_4)\bar{u}_g(k_5))|^2\nn\\
&&{}-2 |\mathcal{M}(b\bar{b}({}^3S_1^{[1]})\rightarrow c\bar{c}({}^3S_1^{[8]}) + u_g(k_3)g\bar{u}_g(k_5))|^2\nn\\
&&{}-\left.2 |\mathcal{M}(b\bar{b}(^3S_1^{[1]})\rightarrow c\bar{c}(^3S_1^{[8]}) + u_g(k_3)\bar{u}_g(k_4)g)|^2\right),
\nonumber\\
&&
\eeqn
where $u_g$ and $\bar{u}_g$ stand for the ghost and antighost, respectively.

In this way, we have to calculate (see Fig.~\ref{RC})
$b\bar{b}({}^3S_1^{[1]})(P)\rightarrow c\bar{c}({}^3S_1^{[8]})(K)
+ g(k_3)g(k_4)g(k_5)$ (96 Feynman diagrams) and
$b\bar{b}({}^3S_1^{[1]})\rightarrow c\bar{c}({}^3S_1^{[8]})
+ gu_g\bar{u}_g/u_g g\bar{u}_g/u_g\bar{u}_gg/q\bar{q}g$ (6 Feynman diagrams 
for each subprocess).
These subprocesses share the same kinematics, for which we define
\beqn
&& s_1 \equiv (P-K)^2,\ s_2\equiv(P-K-k_4)^2,\ u_1\equiv(P-k_4)^2,\nn\\ 
&& u_2\equiv(P-k_5)^2,\ t_2\equiv(P-k_4-k_5)^2.
\eeqn

In the phase space integrations of the above subprocesses, we encounter soft,
collinear, and soft-collinear divergences, which we subtract by means of the
phase space slicing method \cite{Harris:2001sx}.
Specifically, we introduce two slicing parameters, $\delta_s$ and $\delta_c$,
to demarcate the soft regions as 
\beqn
\frac{E_3}{m_c}< \delta_s\ \text{or}\ \frac{E_4}{m_c}< \delta_s\ \text{or}\ \frac{E_5}{m_c}< \delta_s,
\label{softregion}
\eeqn
and the collinear regions as
\beqn
s_{34} < \delta_c m_c^2 \ \text{or}\ s_{35} < \delta_c m_c^2 \ \text{or}\  s_{45} < \delta_c m_c^2,
\label{collinear}
\eeqn 
where $E_3$, $E_4$, and $E_5$ are the energies of the soft gluons with momenta
$k_3$, $k_4$, and $k_5$, respectively, and  
$s_{34}=(k_3+k_4)^2$, $s_{35}=(k_3+k_5)^2$, $s_{45}=(k_4+k_5)^2$.
In our case, where three identical gluons are in the final state,
it is sufficient to subtract the soft and collinear divergences in one of the
soft regions in Eq.~(\ref{softregion}) and one of the  collinear regions in
Eq.~(\ref{collinear}).

\subsubsection{Soft Region}

As an example, let us consider the case when $k_5$ is soft, which implies that
$E_5 < \delta_s m_c$.
We parametrize the momenta in the center-of-mass frame of $P$ and $-K$, which,
at the same time, is the center-of-mass frame of $k_3$ and $k_4$ in the limit
$k_5\rightarrow 0$, as
\beqn
P &=& (E_P,\ 0,\ |\boldsymbol{p}|\sin{\theta},\ |\boldsymbol{p}|\cos{\theta}),
\nonumber\\
K &=& (E_K,\ 0,\ |\boldsymbol{p}|\sin{\theta},\ |\boldsymbol{p}|\cos{\theta}),
\nonumber\\
k_3 &=& E_3(1,\ 0,\ 0,\ 1),
\nonumber\\
k_4 &=& E_4(1,\ 0,\ 0,\ -1),
\nonumber\\
k_5 &=& E_5(1,\ \sin{\theta_1}\sin{\theta_2},\ \sin{\theta_1}\cos{\theta_2},\  \cos{\theta_1}),
\eeqn
where 
\beqn
&& E_P= \frac{s_b+u_b}{2\sqrt{t}},\qquad
E_K= \frac{s_c+u_c}{2\sqrt{t}},\qquad
E_3=E_4= \frac{\sqrt{t}}{2},\nn\\
&&|\boldsymbol{p}|=\frac{a}{2\sqrt{t}},\ \cos{\theta}=\frac{u-s}{a},
\eeqn
with $a=\sqrt{(s+u)^2-64m_c^2m_b^2}$.

The corresponding contribution from the soft region to the real corrections is
given by
\beqn
d\hat{\Gamma}_{k_5\text{soft}}^{\text{RC}}= \frac{1}{4m_b}d \text{PS}_{1\rightarrow 3}\int_{\text{soft}} d \text{PS}_{k_5}\overline{\sum}
|\mathcal{M}_{k_5\text{soft}}|^2,
\eeqn
where 
\beqn
\lefteqn{\int_{\text{soft}} d \text{PS}_{k_5} \equiv \int_{\text{soft}}\frac{\mu^{4-D}d^{D-1} k_5}{2(2\pi)^{D-1}E_5}=\frac{(\pi\mu^2)^\e\Gamma(1-\e)}{(2\pi)^3\Gamma(1-2\e)}} \nn\\
&&{}\times\int_0^{\delta_s m_c}E_5^{1-2\e}dE_5\int_0^\pi \sin{\theta_1}^{1-2\e}d\theta_1 \int_0^\pi \sin{\theta_2}^{-2\e}d\theta_2.\nn\\
\eeqn
Therefore, in the limit $k_5 \rightarrow 0$, we have
\beqn
\lefteqn{d\hat{\Gamma}_{k_5\text{soft}}^{\text{RC}}=d \hat{\Gamma}^{\text{LO}}(b\bar{b}(^3S_1^{[1]})\rightarrow c\bar{c}(^3S_1^{[8]}) + gg)}\nn\\
&&{}\times \frac{3}{2}\int_{\text{soft}} d \text{PS}_{k_5}\left[\frac{t}{(k_3\cdot k_5)(k_4\cdot k_5)} -\frac{8m_c^2}{(K\cdot k_5)^2}\right.\nn\\
&&{}+\left. \frac{u_c}{(k_4\cdot k_5)(K\cdot k_5)}+ \frac{s_c}{(k_3\cdot k_5)(K\cdot k_5)}\right],
\label{softintegral}
\eeqn
where
$\hat{\Gamma}^{\text{LO}}(b\bar{b}({}^3S_1^{[1]})\rightarrow c\bar{c}({}^3S_1^{[8]}) + gg)$ is calculated in $D$ dimensions and the results of the soft integrals
are listed in Appendix~B.

\subsubsection{Hard-Collinear Region}

Let us assume that $k_4$ is collinear to $k_5$.
Then the SDC can be factorized as
\beqn
\lefteqn{d \hat{\Gamma}_{4^\prime\rightarrow 45,\text{hard}}(b\bar{b}({}^3S_1^{[1]})\rightarrow c\bar{c}({}^3S_1^{[8]}) + ggg)}\nn\\
&=& d \hat{\Gamma}^{\text{LO}}(b\bar{b}({}^3S_1^{[1]})\rightarrow c\bar{c}({}^3S_1^{[8]}) + gg)\frac{3g_s^2}{8\pi^2}
\left(\frac{4\pi\mu^2 e^{-\gamma_E}}{m_c^2}\right)^\e\nonumber\\
&&{}\times  \left[\left(2\ln\frac{2\delta_sm_c}{\sqrt{t}}+\frac{11}{6}\right)
  \left(\frac{1}{\e}-\ln{\delta_c}\right)
-\ln^2\frac{2\delta_sm_c}{\sqrt{t}}
  \right.\nonumber\\
  &&{}+\left.
  \frac{67}{18}-\frac{\pi^2}{3}
\right]
\eeqn
for $g\rightarrow gg$ splitting and 
\beqn
\lefteqn{d \hat{\Gamma}_{4^\prime\rightarrow 45,\text{hard}}(b\bar{b}({}^3S_1^{[1]})\rightarrow c\bar{c}({}^3S_1^{[8]}) + g + q\bar{q})} \nn\\
& =& d \hat{\Gamma}^{\text{LO}}(b\bar{b}(^3S_1^{[1]})\rightarrow c\bar{c}(^3S_1^{[8]}) + gg)\nonumber\\
&&{}\times \frac{g_s^2}{24\pi^2}\left(\frac{4\pi\mu^2 e^{-\gamma_E}}{m_c^2}\right)^\e \left(-\frac{1}{\e}+\ln\delta_c-\frac{5}{3}\right)
\eeqn
for $g\rightarrow q\bar{q}$ splitting.

In the case of $g\rightarrow gg$ splitting, hard conditions for the splitting
gluons are applied to avoid double counting of the soft-collinear region.

\subsubsection{Hard-Noncollinear Region}

In the hard-noncollinear region, the phase space integration is finite, so that
we can directly perform the numerical integration in four dimensions.
In the rest frame of $P$ and $-K$, the conditions defining the
hard-noncollinear region are given by
 \beqn
 E_3 &=& \frac{s_1+u_1+u_2-t_2-4m_b^2}{2\sqrt{s_1}} > \delta_s m_c,
\nonumber\\
 E_4 &=& \frac{s_1-s_2}{2\sqrt{s_1}} > \delta_s m_c,
\nonumber\\
 E_5 &=& \frac{t_2+s_2-u_1-u_2+4m_b^2}{2\sqrt{s_1}} > \delta_s m_c,
\nonumber\\
 s_{34} &=& s_1+u_1+u_2-t_2-s_2-4m_b^2 > \delta_c m_c^2,
\nonumber\\
 s_{35} &=& s_2 > \delta_c m_c^2,
\nonumber\\
 s_{45} &=& 4m_b^2+t_2-u_1-u_2 > \delta_c m_c^2.
 \eeqn
 We express the four-body phase space in covariant form \cite{Kumar:1970cr}, so
 that these conditions can be easily implemented in the numerical phase space
 integration.
 Combining the contributions from the soft, hard-collinear, and
 hard-noncollinear regions, the numerical results converge as $\delta_s$
 becomes small ($\delta_s<10^{-2}$) with $\delta_c$ being much smaller than
 $\delta_s$.
 Specifically, we choose $\delta_c=\delta_s/100$.
  
\subsection{$b\bar{b}({}^3S_1^{[1]})\rightarrow c\bar{c}({}^3P_J^{[1]}) + ggg$}
\begin{figure}
\begin{center}
\includegraphics[width=0.9\linewidth]{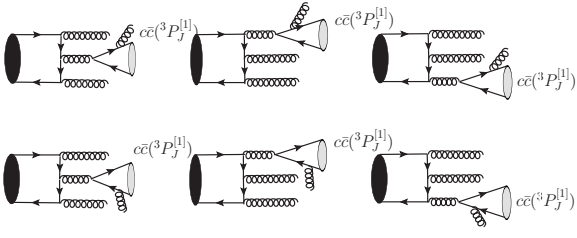}
\caption{Typical tree-level Feynman diagrams for the partonic subprocess
  $b\bar{b}({}^3S_1^{[1]})\rightarrow c\bar{c}({}^3P_J^{[1]}) + ggg$.}
\label{gggpwave}
\end{center}
\end{figure}

There are 36 Feynman diagrams for the tree-level subprocess
$b\bar{b}({}^3S_1^{[1]})\rightarrow c\bar{c}({}^3P_J^{[1]}) + ggg$ (see
Fig.~\ref{gggpwave}).
As expected, IR divergences appear in the phase space integration.
In NRQCD factorization, this kind of IR divergences are absorbed into the NLO
corrections to the respective CO LDME
$\langle{\cal O}^{\chi_{cJ}}({}^3S_1^{[8]})\rangle$
\cite{Bodwin:1994jh,Butenschon:2009zza},
\beqn
\lefteqn{\langle{\cal O}^{\chi_{cJ}}({}^3S_1^{[8]})\rangle_{\text{Born}}
  = \langle{\cal O}^{\chi_{cJ}}({}^3S_1^{[8]})\rangle_{\text{ren}}}\nn\\
&&{}+\frac{2\alpha_s}{3\pi m_c^2}
\left(\frac{4\pi\mu^2}{\mu^2_\Lambda}e^{-\gamma_E}\right)^\epsilon
\frac{1}{\epsilon_{\text{IR}}}\sum_J\left[\frac{C_F}{C_A}
  \langle{\cal O}^{\chi_{cJ}}({}^3P_J^{[1]})\rangle_{\text{Born}}\right.\nn\\
  &&{}+\left.\left(\frac{C_A}{2}-\frac{2}{C_A}\right)
  \langle{\cal O}^{\chi_{cJ}}({}^3P_J^{[8]})\rangle_{\text{Born}}\right],
\eeqn
where $\langle{\cal O}^{\chi_{cJ}}({}^3S_1^{[8]})\rangle_{\text{ren}}$ is the
renormalized LDME and $\mu_\Lambda$ is the renormalization scale of the LDME 
$\langle\mathcal{O}^{\chi_{cJ}}({}^3S_1^{[8]})\rangle$.

Here, we adopt the same approach as in the computation of the real corrections
to $b\bar{b}({}^3S_1^{[1]})\rightarrow c\bar{c}({}^3S_1^{[8]}) + ggg$ to extract
the IR divergences, except that only one slicing parameter, namely $\delta_s$,
is needed.
Let us consider the limit of $k_5$ being soft as an example, in which the
squared matrix element can be factorized as 
\beqn
\lefteqn{|\mathcal{M}(b\bar{b}({}^3S_1^{[1]})\rightarrow c\bar{c}({}^3P_J^{[1]})
  + ggg)|^2_{k_5\text{soft}}}\nn\\
 &=& 4g_s^2\frac{\epsilon^{\beta^\prime}(k_5)\epsilon^{*\beta}(k_5)\varepsilon_{\alpha\beta}^{(J)*}(K)\varepsilon_{\alpha^\prime\beta^\prime}^{(J)}(K)}{(K\cdot k_5)^2}\nn\\
 &&{}\times \mathcal{M}_{\text{Born}}^\alpha(T_c-T_{\bar{c}})(T_c-T_{\bar{c}})\mathcal{M}^{*\alpha^\prime}_{\text{Born}},
 \label{softlimit}
\eeqn
where $\epsilon^{*\beta}(k_5)$ and $\varepsilon_{\alpha\beta}^{(J)*}(K)$ are the
polarization vector and tensor of the soft gluon and $c\bar{c}$ pair,
$T_c$ and $T_{\bar{c}}$ are the color matrices corresponding to the soft-gluon
attachments to the charm and anticharm quark lines, and
$\mathcal{M}_{\text{Born}}^\alpha$ is the amplitude of the Born-level subprocess
$b\bar{b}({}^3S_1^{[1]})\rightarrow c\bar{c}({}^3S_1^{[8]}) + gg$, with $\alpha$
being the Lorentz index of the polarization vector of the $c\bar{c}$ pair.

Choosing axial gauge for the soft gluon, with polarization sum
\beqn
\lefteqn{\sum_{\text{pol}}\epsilon_{\beta^\prime}(k_5)\epsilon_{\beta}^*(k_5)}\nn\\
&=&-g_{\beta^\prime\beta} + \frac{K_{\beta^\prime} k_{5\beta}+K_\beta k_{5\beta^\prime}}{K\cdot k_5} -\frac{K^2k_{5\beta^\prime}k_{5\beta}}{(K\cdot k_5)^2},
\eeqn
the contribution from the region where $k_5$ is soft then reads
\beqn
d\hat{\Gamma}_{k_5\text{soft}}&=& \frac{1}{4m_b}d \text{PS}_{1\rightarrow 3}\int_{\text{soft}} d \text{PS}_{k_5}\nn\\
&&{}\times\overline{\sum}
|\mathcal{M}(b\bar{b}({}^3S_1^{[1]})\rightarrow c\bar{c}({}^3P_J^{[1]}) + ggg)|^2_{k_5\text{soft}}\nn\\
&=& \frac{1}{4m_b}d \text{PS}_{1\rightarrow 3}\int_{\text{soft}} d \text{PS}_{k_5}\frac{4g_s^2}{(K\cdot k_5)^2}\nn\\
&&\times
\left(-g_{\beta^\prime\beta} + \frac{K_\beta^\prime k_{5\beta}+K_\beta k_{5\beta^\prime}}{K\cdot k_5} -\frac{K^2k_{5\beta^\prime}k_{5\beta}}{(K\cdot k_5)^2}\right)\nn\\
&&{}\times\overline{\sum}\varepsilon_{\alpha\beta}^{(J)*}\varepsilon_{\alpha^\prime\beta^\prime}^{(J)} \mathcal{M}_{\text{Born}}^\alpha(T_c-T_{\bar{c}})(T_c-T_{\bar{c}})
\nn\\
&&{}\times\mathcal{M}^{*\alpha^\prime}_{\text{Born}}.
\eeqn

Now we tackle the new type of tensor integral
$\int_{\text{soft}} d \text{PS}_{k_5} \frac{k_{5\beta}k_{5\beta^\prime}}{(K\cdot k_5)^4}$.
This cannot be reduced to scalar integrals through conventional tensor
reduction procedures, since it is not Lorentz covariant due to the cut-off in
the soft-gluon energy. 
Therefore, we explicitly evaluate these tensor integrals as they appear, e.g.\
as $\int_{\text{soft}} d \text{PS}_{k_5} \frac{(k_3\cdot k_5)(k_4\cdot k_5)}{(K\cdot k_5)^4}$.
As a result, the $D$-dimensional Born-level squared matrix element of
subprocess $b\bar{b}({}^3S_1^{[1]})\rightarrow c\bar{c}({}^3S_1^{[8]}) + gg$
does not factorize from the $D$-dimensional squared matrix elements of
subprocess
$b\bar{b}({}^3S_1^{[1]})\rightarrow c\bar{c}({}^3P_J^{[1]}) + ggg$ in the soft
limit.
This is different from the case of the real corrections.
Nevertheless, this does not affect the cancellation of IR divergences, since
the divergences appear to be proportional to the four-dimensional Born-level
squared matrix element of subprocess
$b\bar{b}({}^3S_1^{[1]})\rightarrow c\bar{c}({}^3S_1^{[8]}) + gg$.

\section{Phenomenological Results}

\begin{figure*}
\centering
\begin{tabular}{cc}
\includegraphics[width=0.45\linewidth]{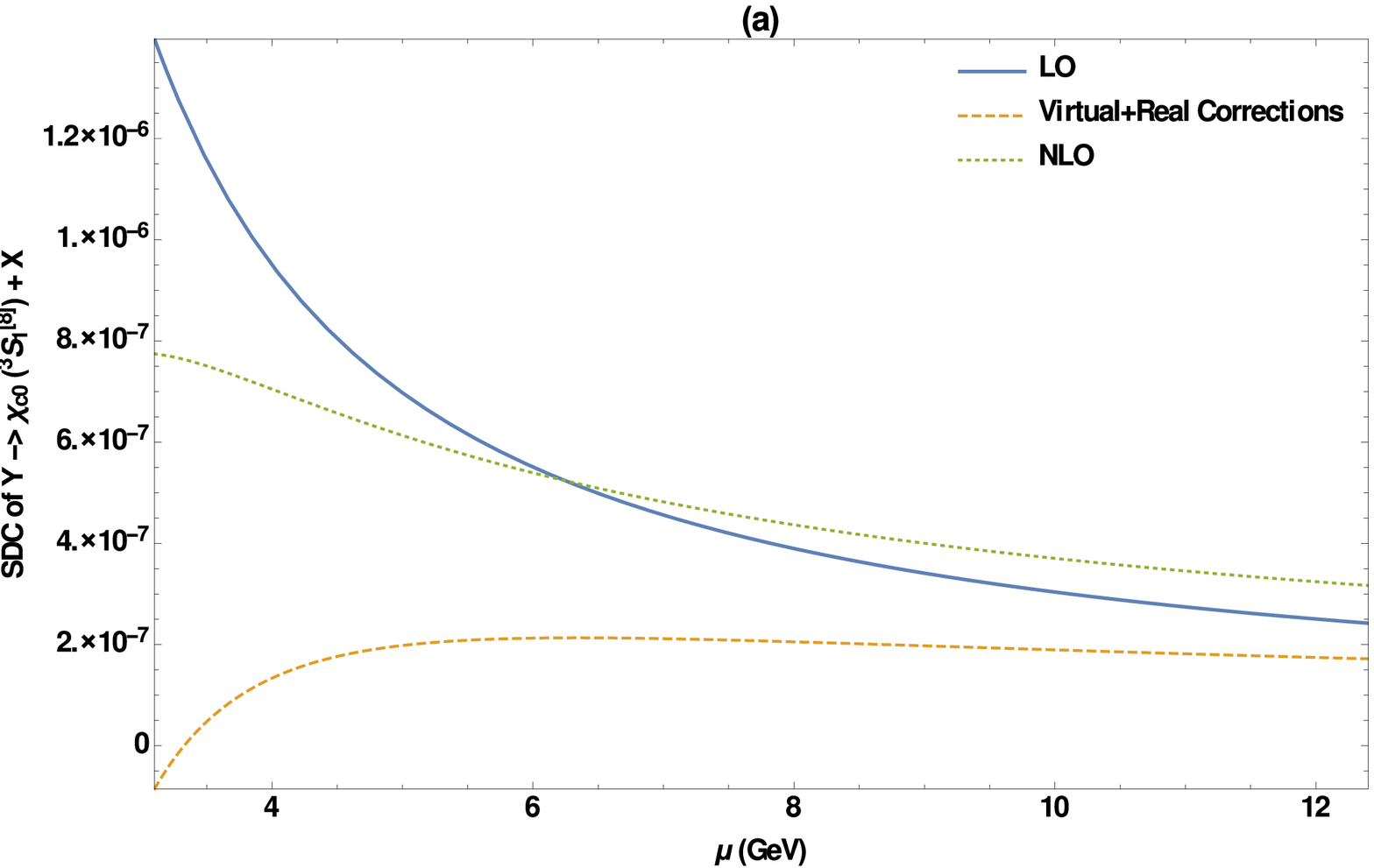} &
\includegraphics[width=0.422\linewidth]{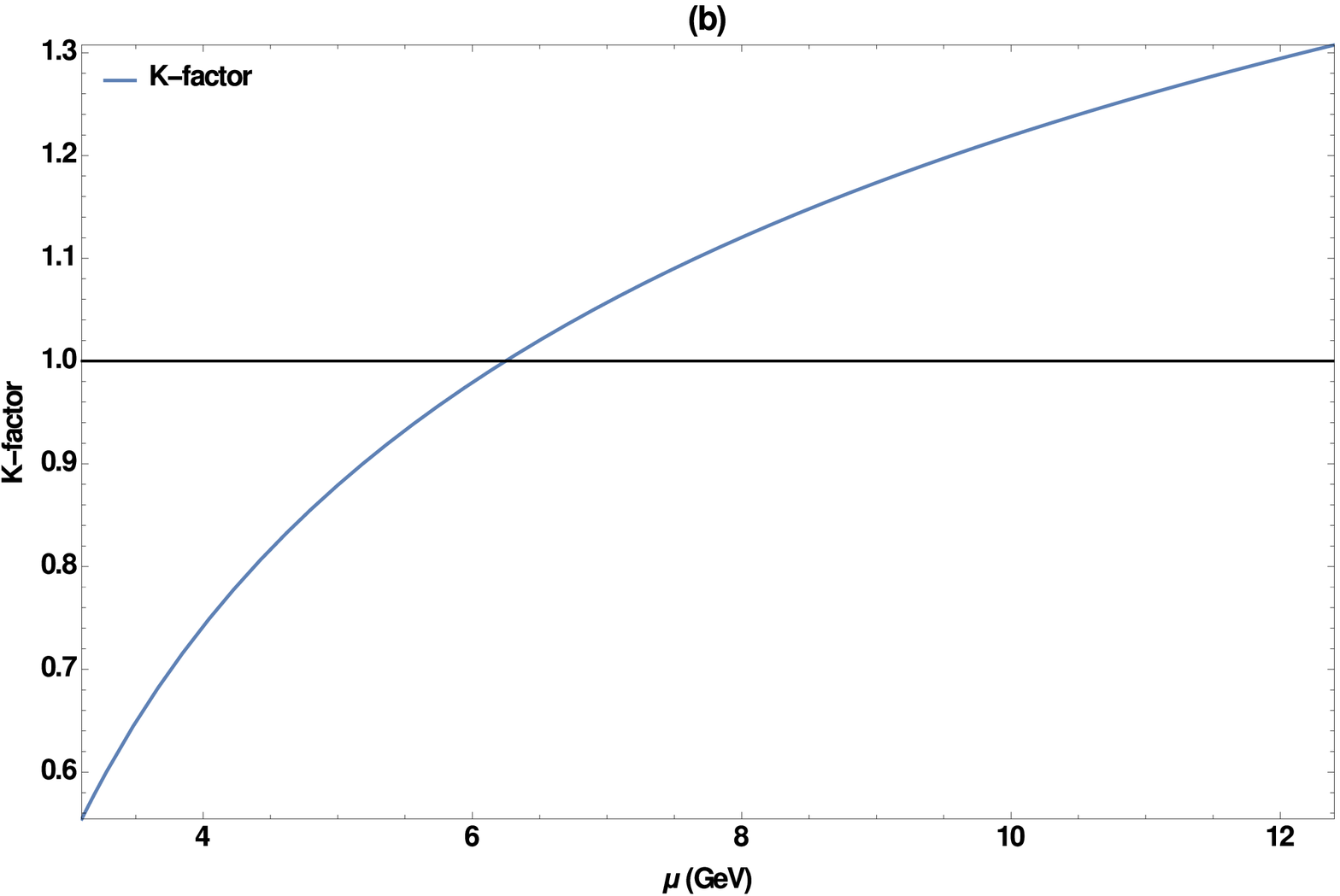}
\end{tabular}
\caption{$\mu$ dependencies of
  (a) $\hat{\Gamma}_{8,gg}^{\text{LO}}$, $\hat{\Gamma}_{8,gg}^{\text{corr}}$,
 and $\hat{\Gamma}_{8,gg}^{\text{NLO}}$ and (b) $K$ factor.}
\label{COSDC}
\end{figure*}

We are now in the position to present our numerical analysis of inclusive
$\chi_{cJ}$ production in $\Upsilon$ decay at $\mathcal{O}(\alpha_s^5)$ in the
NRQCD factorization framework.
We first list our input parameters.
We take the quark pole masses to be $m_c=m_{J/\psi}/2= 1.5$~GeV and 
$m_b=m_{\Upsilon}/2=4.75$~GeV, and fix the NRQCD factorization scale to be
$\mu_\Lambda=m_c$.
For consistency, we employ the one-loop (two-loop) formula for
$\alpha_s(\mu)$ \cite{Patrignani:2016xqp} in the LO (NLO) part of our analysis,
with $n_f=3$ active quark flavors and asymptotic scale parameter
$\Lambda_{\text{QCD}}=249$~MeV (389~MeV).
Adopting the value $|R^{\prime}_{1P}(0)|=0.075~\mathrm{GeV}^5$ of the first 
derivative of the wave function at the origin obtained in
Ref.~\cite{Eichten:1995ch} for the Buchm\"uller-Tye potential
\cite{Buchmuller:1980su}, we have
$\langle\mathcal{O}^{\chi_{c0}}({}^3P_0^{[1]})\rangle=0.107$~GeV$^5$.
As for the CO LDME $\langle\mathcal{O}^{\chi_{c0}}({}^3S_1^{[8]})\rangle$, we
use the results 
$(2.21\pm0.12)\times10^{-3}~\mathrm{GeV}^{3}$ \cite{Gong:2012ug} and
$(2.2^{+0.5}_{-0.3})\times10^{-3}~\mathrm{GeV}^{3}$ \cite{Ma:2010vd}
extracted by fitting to experimental data of $\chi_{cJ}$ hadroproduction.

We start by investigating the sizes and $\mu$ dependencies of the SDCs of the
individual partonic subprocesses.
It is convenient to write:
\begin{eqnarray}
\hat{\Gamma}_{8,gg}^{\text{LO}}&=&f_{8,gg}^{\text{LO}}(m_c,m_b)\alpha_s^4(\mu)~\text{GeV}^{-5},
\nonumber\\
\hat{\Gamma}_{8,gg}^{\text{corr}}&=&f_{8,gg}^{\text{corr}}(m_c,m_b,\mu)\alpha_s^5(\mu)~\text{GeV}^{-5},
\nonumber\\
\hat{\Gamma}_{8,c\bar{c}g}&=&f_{8,c\bar{c}g}(m_c,m_b)\alpha_s^5(\mu)~\text{GeV}^{-5},
\nonumber\\
\hat{\Gamma}_{1,ggg}^J&=&f_{1,ggg}^J(m_c,m_b)\alpha_s^5(\mu)~\text{GeV}^{-7},\nonumber\\
\hat{\Gamma}_{1,c\bar{c}g}^J&=&f_{1,c\bar{c}g}^J(m_c,m_b)\alpha_s^5(\mu)\text{GeV}^{-7},
\end{eqnarray}
where $\hat{\Gamma}_{1/8,X}^J$ is the SDC of partonic subprocess
$\Upsilon({}^3S_1^{[1]})\to \chi_{cJ}({}^3P_J^{[1]}/{}^3S_1^{[8]})+X$, and the
alternative labels "LO" and "corr" stand for the LO contribution and the
radiative correction to it.
Our numerical results for the dimensionless factors $f_{1/8,X}$ are listed in
Table~\ref{table:1}.
From there, we observe that there are strong numerical cancellations between
the $c\bar{c}({}^3P_J^{[1]})$ and $c\bar{c}({}^3S_1^{[8]})$ channels.
In fact, the entries for $f_{1,ggg}^J$ in Table~\ref{table:1} are negative and
do not carry any physical meaning by themselves.
This may be understood by observing that the CS SDCs of
$b\bar{b}({}^3S_1^{[1]})\rightarrow c\bar{c}({}^3P_J^{[1]}) + ggg$ are IR
divergent to start with \cite{Trottier:1993ze}.
Therefore, $\Upsilon\to\chi_{cJ}+X$ as a strictly inclusive decay falls outside
the range of applicability of the CS model \cite{Berger:1980ni,Baier:1981uk}.
Within NRQCD factorization, these CS SDCs are rendered finite by absorbing
their IR divergences into the CO LDME
$\langle\mathcal{O}^{\chi_{c0}}({}^3S_1^{[8]})\rangle$ associated with the
SDC of $b\bar{b}({}^3S_1^{[1]})\rightarrow c\bar{c}({}^3S_1^{[8]}) + gg$ at LO.
Consequently, only the combinations of the $c\bar{c}({}^3P_J^{[1]})$ and
$c\bar{c}({}^3S_1^{[8]})$ channels may be interpreted as physical
observables.
This mechanism of IR cancellation by NRQCD factorization is familiar, e.g.,
from hadronic $h_c$ decay \cite{Huang:1996fa}.
Furthermore, Table~\ref{table:1} tells us that the contributions from
$\chi_{cJ}$ production in association with open charm are greatly suppressed by
phase space as expected.
\begin{table}[h!]
\begin{tabular}{c|c|c}
\hline\hline
$f_{8,gg}^{\text{LO}}(10^{-4})$ & $f_{8,gg}^{\text{corr}}(10^{-4})$ &
$f_{8,c\bar{c}g}(10^{-5})$\\
\hline
$2.38$ & $4.85 +13.62\ln(\frac{\mu}{m_b})$& $1.23$ \\
\hline\hline
 $f_{1,ggg}^0(10^{-5})$& $f_{1,ggg}^1(10^{-5})$& $f_{1,ggg}^2(10^{-5})$\\
\hline
 $-4.18$& $-2.06$ & $-2.65$\\
\hline\hline
 $f_{1,c\bar{c}g}^0(10^{-7})$& $f_{1,c\bar{c}g}^1(10^{-7})$ &
$f_{1,c\bar{c}g}^2(10^{-7})$\\
\hline
 $1.73$& $1.04$ & $0.35$\\
\hline\hline
\end{tabular}
\caption{Numerical values of $f_{8,gg/c\bar{c}g}^{\text{LO}}$,
$f_{8,gg}^{\text{corr}}$, and $f_{1,ggg/c\bar{c}g}^J$.}
\label{table:1}
\end{table}

In Fig.~\ref{COSDC}(a), we study the dependencies on the renormalization scale
$\mu$ of $\hat{\Gamma}_{8,gg}^{\text{LO}}$, $\hat{\Gamma}_{8,gg}^{\text{corr}}$, and
$\hat{\Gamma}_{8,gg}^{\text{NLO}}=\hat{\Gamma}_{8,gg}^{\text{LO}}
+\hat{\Gamma}_{8,gg}^{\text{corr}}$.
As mentioned above, $\hat{\Gamma}_{8,gg}^{\text{LO}}$ is evaluated with the
one-loop expression of $\alpha_s(\mu)$, while $\hat{\Gamma}_{8,gg}^{\text{corr}}$ and
$\hat{\Gamma}_{8,gg}^{\text{NLO}}$ are evaluated with the two-loop expression of
$\alpha_s(\mu)$.
In Fig.~\ref{COSDC}(b), the factor
$K=\hat{\Gamma}_{8,gg}^{\text{NLO}}/\hat{\Gamma}_{8,gg}^{\text{LO}}$ is shown as a
function of $\mu$.
As expected on general grounds, the $\mu$ dependence is reduced as we pass from
LO to NLO.
Unfortunately, the $\mu$ dependence of $\hat{\Gamma}_{8,gg}^{\text{NLO}}$ is still
appreciable, so that scale optimization appears appropriate.
Since the $\mu$ dependence of $\hat{\Gamma}_{8,gg}^{\text{NLO}}$ is monotonic, the
Principle of Minimal Sensitivity \cite{Stevenson:1980du} is not applicable.
However, the concept of Fastest Apparent Convergence (FAC)
\cite{Grunberg:1980ja} works, since the value of $\mu$ for which $K=1$,
$\mu_{\text{FAC}}$, is a typical energy scale of the $\Upsilon\to\chi_{cJ}+X$
decays.
From Fig.~\ref{COSDC}, we read off that $\mu_{\text{FAC}}=6.2$~GeV.
In the following, we will use this as the central scale and estimate the
theoretical uncertainty by varying $\mu$ in the range
$\mu_{\text{FAC}}/2 <\mu<2\mu_{\text{FAC}}$.

\begin{figure*}
\centering
\begin{tabular}{cc}
\includegraphics[width=0.45\linewidth]{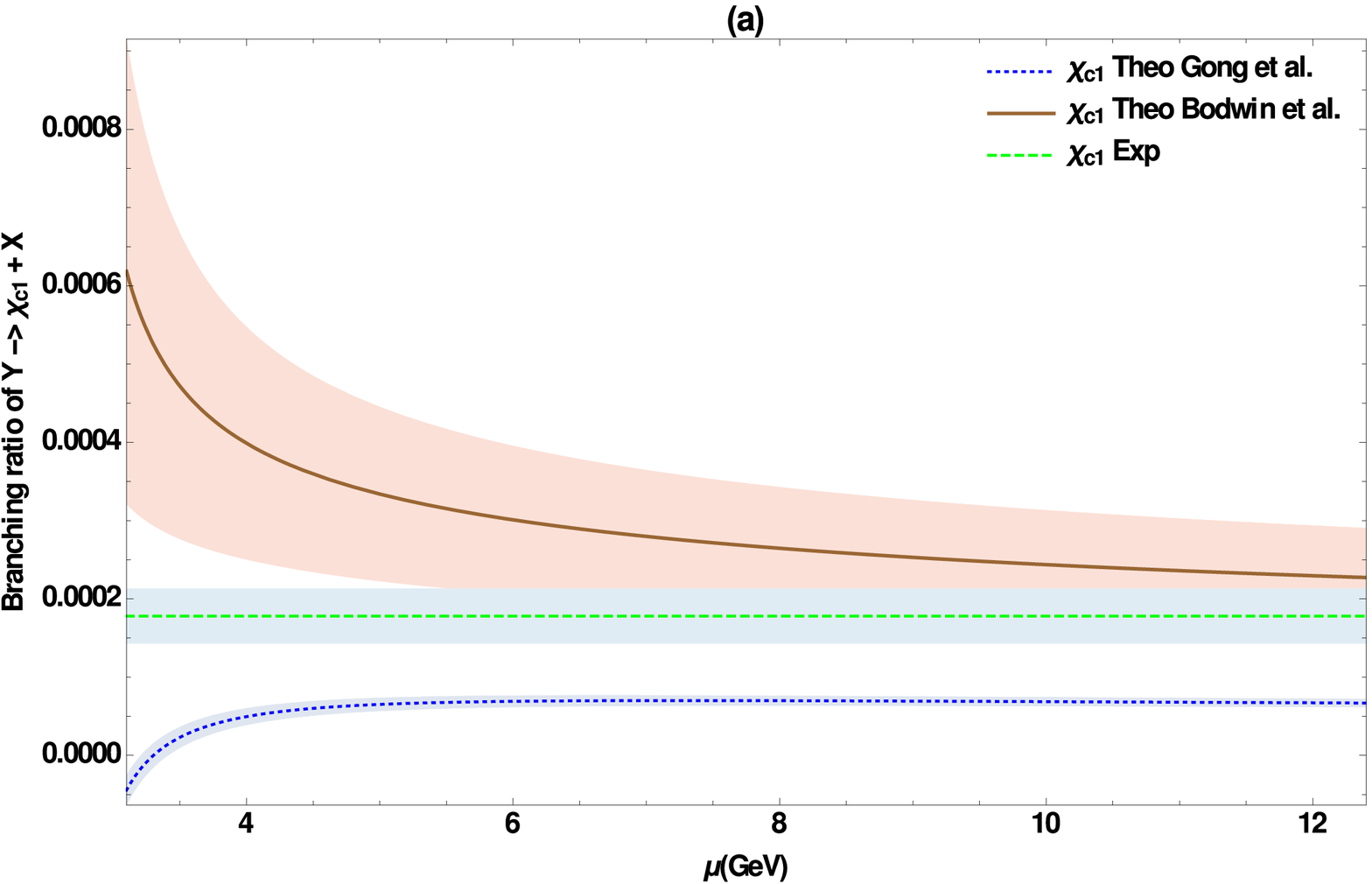} &
\includegraphics[width=0.45\linewidth]{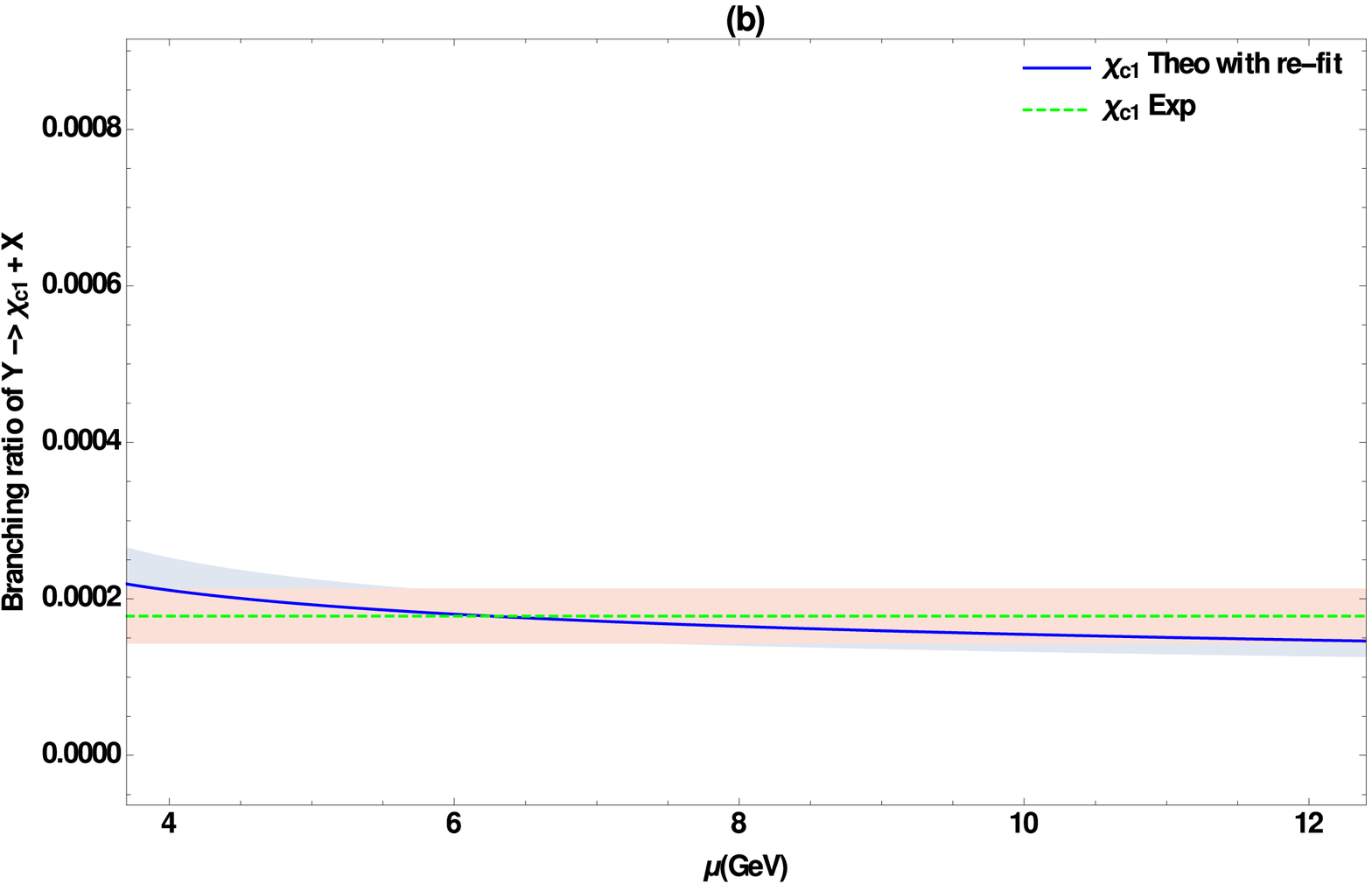} \\
\includegraphics[width=0.45\linewidth]{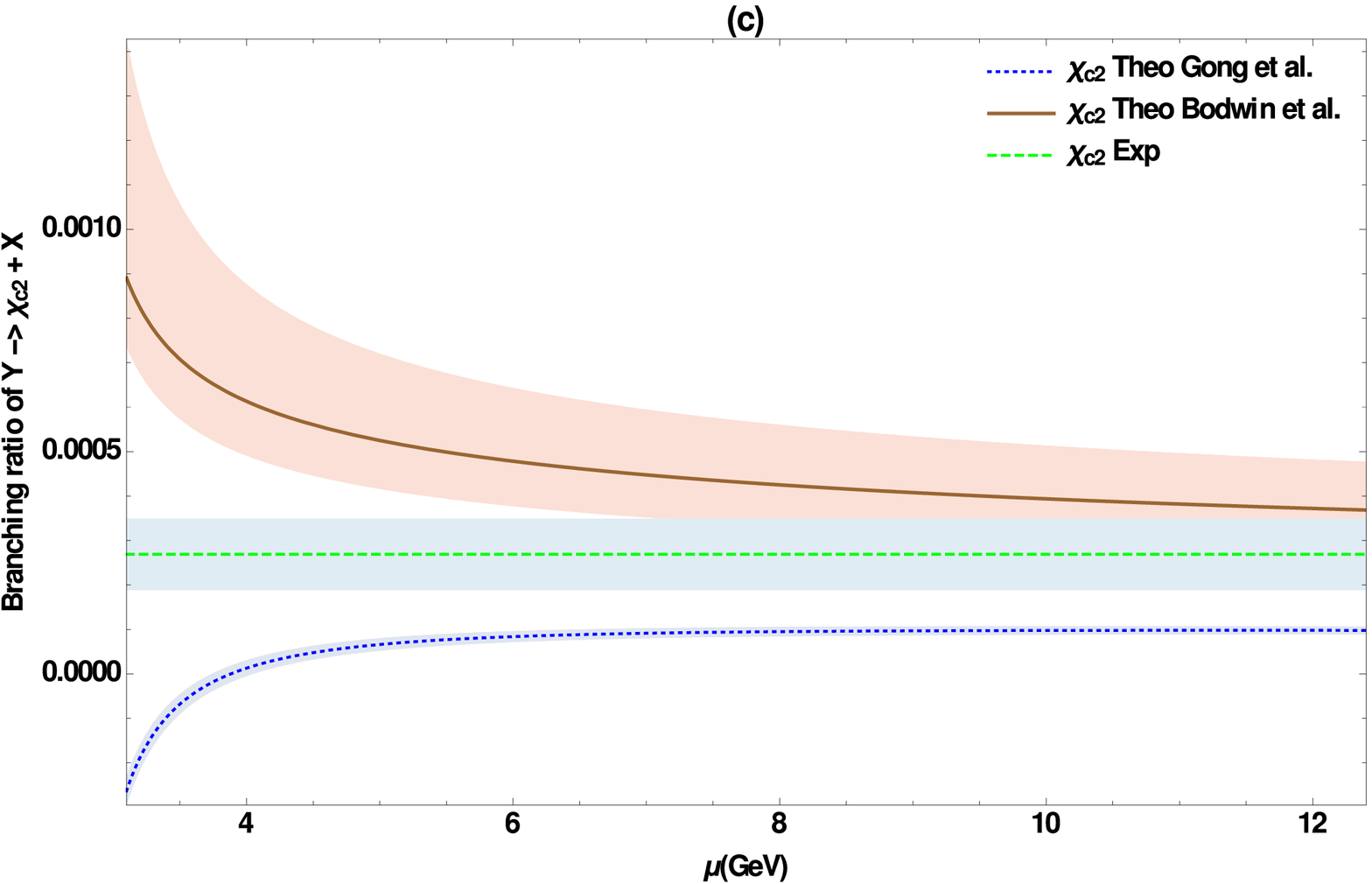} &
\includegraphics[width=0.45\linewidth]{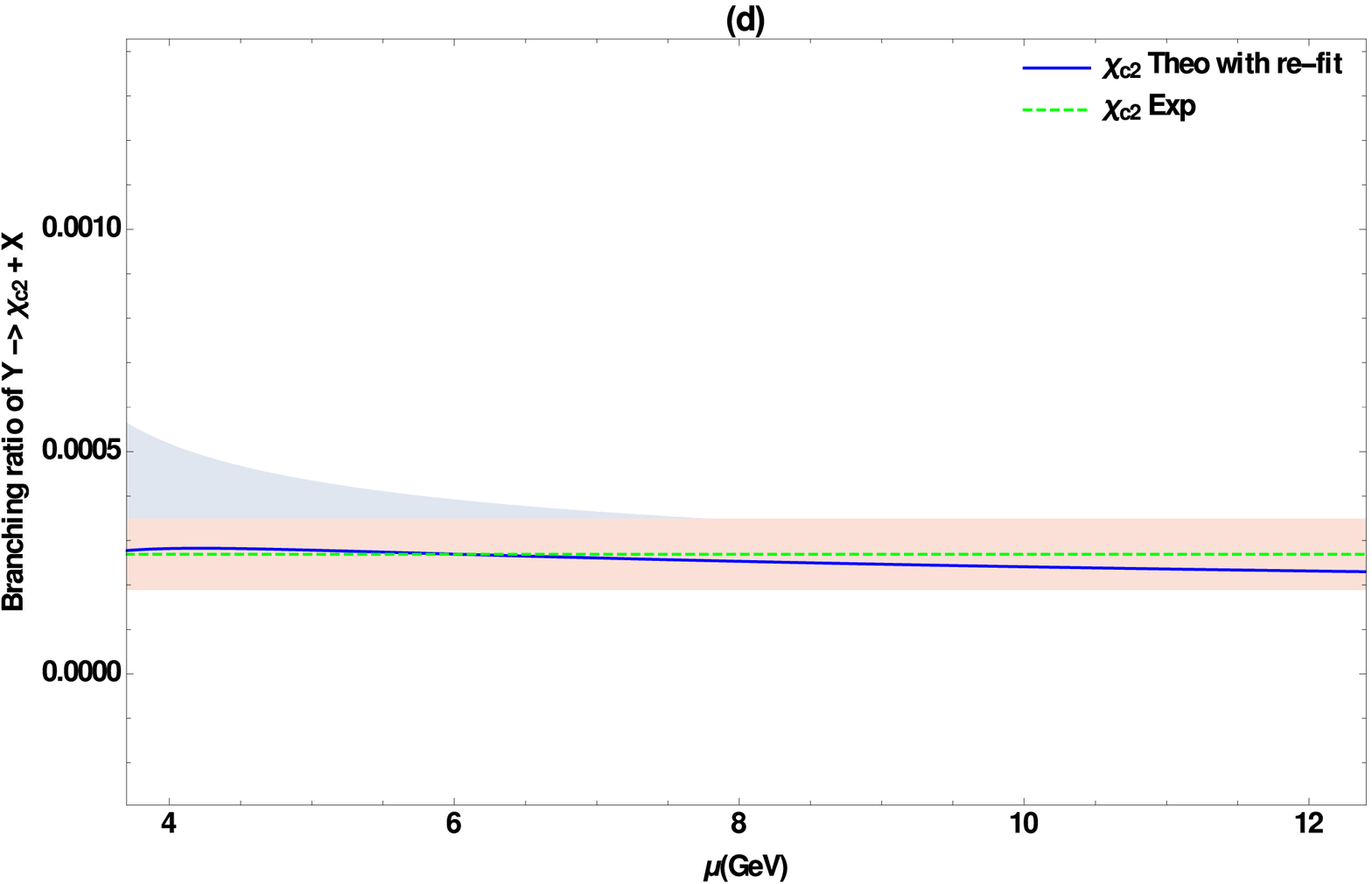}
\end{tabular}
\caption{$\mu$ dependencies of $\mathcal{B}(\Upsilon\to \chi_{c1}+X)$
(upper row) and $\mathcal{B}(\Upsilon\to \chi_{c2}+X)$ (lower row) evaluated
with different choices of LDMEs, to be compared with experimental values
$\mathcal{B}(\Upsilon\to \chi_{c1}+X)=(1.78\pm0.35)\times 10^{-4}$ and
$\mathcal{B}(\Upsilon\to \chi_{c2}+X)=(2.69\pm 0.8)\times 10^{-4}$
\cite{Patrignani:2016xqp}, respectively.}
\label{Brhadron}
\end{figure*}

From Figs.~\ref{ggtree}--\ref{gggpwave}, we observe that $\chi_{cJ}$
production from $\Upsilon$ decay may be viewed, at least at LO, as
$\Upsilon\to ggg$ decay followed by $g\to \chi_{cJ}+X$ via fragmentation.
Inspired by this observation, we express the branching ratio
$\mathcal{B}(\Upsilon \rightarrow \chi_{cJ} + X)$ as
\cite{He:2009by,He:2010cb}
\beqa
\mathcal{B}(\Upsilon \rightarrow \chi_{cJ} + X)
=\frac{\Gamma (\Upsilon \rightarrow \chi_{cJ} + X)}
{\Gamma(\Upsilon \rightarrow ggg)}
\mathcal{B}(\Upsilon \rightarrow ggg)\nn.\\
\eeqa
We evaluate the partial decay width $\Gamma(\Upsilon \rightarrow ggg)$ through
 $\mathcal{O}(\alpha_s^4)$ as \cite{Mackenzie:1981sf}
\beqn
\lefteqn{\Gamma(\Upsilon \rightarrow ggg) = \frac{20\alpha_s^{ 3}(\mu)}{243m_b^2}(\pi^2-9)\langle \Upsilon | \mathcal{O}(^3S_1^{[1]}) |\Upsilon\rangle}\nn\\
&&{}\times \left\{1+\frac{\alpha_s(\mu)}{\pi}\left[-19.4+
  \frac{3\beta_0}{2}\left(1.161+\ln{(\frac{\mu}{m_b})}\right)\right]\right\},
\nn\\
\eeqn 
and use $\mathcal{B}(\Upsilon \rightarrow ggg)= 81.7 \%$ as determined by the
Particle Data Group \cite{Patrignani:2016xqp}.
This has the advantage that the theoretical prediction no longer depends on
$\langle \Upsilon | \mathcal{O}({}^3S_1^{[1]}) |\Upsilon\rangle$ and that its
dependencies on $\alpha_s$ and $m_b$ are significantly suppressed, so that the
parametric uncertainty is greatly reduced.
By the same token, the $\mu$ dependence is greatly reduced as well to become
quite moderate, as may be seen from Fig.~\ref{Brhadron} to be discussed below.

In the following, we compare our theoretical predictions thus improved with
the experimental results for direct production,
$\mathcal{B}(\Upsilon\to \chi_{c1}+X)=(1.78\pm0.35)\times 10^{-4}$ and
$\mathcal{B}(\Upsilon\to \chi_{c2}+X)=(2.69\pm 0.8)\times 10^{-4}$, obtained
from Ref.~\cite{Patrignani:2016xqp} after subtracting the contributions due to
the feed-down from $\psi(2S)$ mesons.
If we adopt the value
$\langle\mathcal{O}^{\chi_{c0}}({}^3S_1^{[8]})\rangle
=(2.21\pm0.12)\times10^{-3}~\mathrm{GeV}^{3}$ from
Ref.~\cite{Gong:2012ug}, then our theoretical predictions for
$\mathcal{B}(\Upsilon\to \chi_{c1}+X)$ and 
$\mathcal{B}(\Upsilon\to \chi_{c2}+X)$, whose $\mu$ dependencies are displayed
in Figs.~\ref{Brhadron}(a) and (c), respectively, undershoot the experimental
data for all values of $\mu$ in the ballpark of $\mu_{\text{FAC}}$. 

\begin{figure}
\begin{center}
\includegraphics[width=0.9\linewidth]{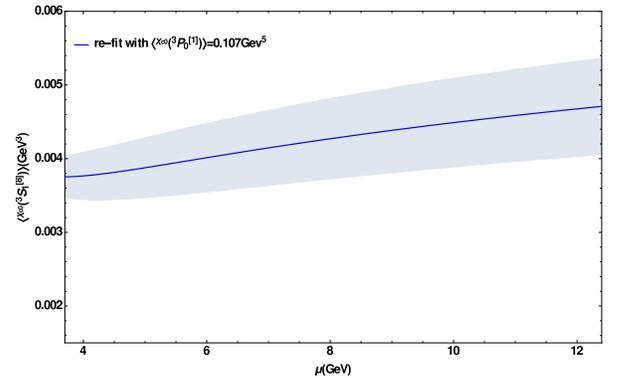}
\caption{$\mu$ dependence of fit value of
$\langle\mathcal{O}^{\chi_{c0}}({}^3S_1^{[8]})\rangle$ for
$\langle\mathcal{O}^{\chi_{c0}}({}^3P_0^{[1]})\rangle = 0.107$~GeV$^5$.}
\label{fit}
\end{center}
\end{figure}

It is interesting to find out which value of
$\langle\mathcal{O}^{\chi_{c0}}({}^3S_1^{[8]})\rangle$ is favored by the
experimental data on $\mathcal{B}(\Upsilon\to \chi_{c1}+X)$ and
$\mathcal{B}(\Upsilon\to \chi_{c2}+X)$.
We, therefore, perform a fit to these data, with the result that
\begin{equation}
\langle\mathcal{O}^{\chi_{c0}}({}^3S_1^{[8]})\rangle
=(4.04\pm0.47)\times10^{-3}~\text{GeV}^3,
\label{fitco}
\end{equation}
for $\langle\mathcal{O}^{\chi_{c0}}({}^3P_0^{[1]})\rangle = 0.107$~GeV$^5$ and our
default choice $\mu=\mu_{\text{FAC}}$.
As for the central values, this is about twice as large as the results of
Refs.~\cite{Ma:2010vd,Gong:2012ug}.
From Figs.~\ref{Brhadron}(b) and (d), we observe that Eq.~(\ref{fitco}) yields
an excellent description of the measurements of
$\mathcal{B}(\Upsilon\to \chi_{c1}+X)$ and
$\mathcal{B}(\Upsilon\to \chi_{c2}+X)$, respectively, throughout the considered
$\mu$ range, $3.7~\text{GeV}<\mu<2\mu_{\text{FAC}}$.
The lower bound on $\mu$ is to ensure the positivity of
$\mathcal{B}(\Upsilon\to \chi_{c0}+X)$.
Thanks to cancellations between the ${}^3P_J^{[1]}$ and ${}^3S_1^{[8]}$
channels, the $\mu$ dependencies are relatively mild indicating small
theoretical uncertainties.

Vice versa, this feature may be exploited to argue that the theoretical
uncertainty in $\langle\mathcal{O}^{\chi_{c0}}({}^3S_1^{[8]})\rangle$ is small.
To this end, we repeat our fit for different values of $\mu$ and show the
outcome in Fig.~\ref{fit}.
We read off from Fig.~\ref{fit} that
$\langle\mathcal{O}^{\chi_{c0}}({}^3S_1^{[8]})\rangle$ ranges from
$(3.7\pm0.28)\times10^{-3}$~GeV$^3$ at $\mu=3.7$~GeV to
$(4.71\pm0.65)\times10^{-3}$~GeV$^3$ at $\mu=2\mu_{\text{FAC}}$.
Obviously, the theoretical uncertainty is comparable to the experimental one.

In Ref.~\cite{Bodwin:2015iua}, an alternative set of $\chi_{cJ}$ production
LDMEs was obtained by fitting cross sections of prompt $\chi_{c1}$ and
$\chi_{c2}$ hadroproduction measured by ATLAS \cite{ATLAS:2014ala} using NLO
SDCs in combination with leading-power fragmentation functions:
\begin{eqnarray}
  \frac{\langle\mathcal{O}^{\chi_{c0}}({}^3P_0^{[1]})\rangle}{m_c^2} &=&
  (3.53\pm 1.08)\times 10^{-2}~\text{GeV}^3,
\nonumber\\
  \langle\mathcal{O}^{\chi_{c0}}({}^3S_1^{[8]})\rangle &=&
  (5.74\pm 1.31)\times 10^{-3}~\text{GeV}^3.
\end{eqnarray}
The resulting predictions for $\mathcal{B}(\Upsilon\to \chi_{c1}+X)$ and
$\mathcal{B}(\Upsilon\to \chi_{c2}+X)$, which are also included in
Figs.~\ref{Brhadron}(a) and (c), respectively, overshoot the experimental data
for $\mu<\mu_{\text{FAC}}$, but the error bands overlap for
$\mu\agt\mu_{\text{FAC}}$.

\section{Summary}

In this work, we studied the inclusive decays $\Upsilon\to\chi_{cJ}+X$ through
$\mathcal{O}(\alpha_s^5)$ and to LO in $v_c$ and $v_b$ in the
NRQCD \cite{Caswell:1985ui} factorization approach \cite{Bodwin:1994jh}.
According to the velocity scaling rules \cite{Lepage:1992tx}, we thus included
the Fock states $b\bar{b}({}^3S_1^{[1]})$, $c\bar{c}({}^3P_J^{[1]})$, and
$c\bar{c}({}^3S_1^{[8]})$.
Besides gluons and light quarks, we also allowed for open charm to appear in
the hadronic debris $X$.
Since the partonic subprocess
$b\bar{b}({}^3S_1^{[1]})\to c\bar{c}({}^3S_1^{[8]})+gg$ already contributes at
$\mathcal{O}(\alpha_s^4)$, we calculated its quantum corrections at NLO.
Since the dependencies of the branching ratios
$\mathcal{B}(\Upsilon\to\chi_{cJ}+X)$ on the renormalization scale $\mu$ turned
out to be monotonic, we applied scale optimization via FAC
\cite{Grunberg:1980ja}, which led to $\mu_{\text{FAC}}=6.2$~GeV.
We then estimated the theoretical uncertainty by varying $\mu$ in the range
between $\mu_{\text{FAC}}/2$ and $2\mu_{\text{FAC}}$.
Using as input the value
$\langle\mathcal{O}^{\chi_{c0}}({}^3P_0^{[1]})\rangle=0.107$~GeV$^5$ obtained in
Ref.~\cite{Eichten:1995ch} for the Buchm\"uller-Tye potential
\cite{Buchmuller:1980su}, we fitted $\langle\mathcal{O}^{\chi_{c0}}({}^3S_1^{[8]})$
to experimental data of $\mathcal{B}(\Upsilon\to \chi_{c1}+X)$ and
$\mathcal{B}(\Upsilon\to \chi_{c2}+X)$ via direct production
\cite{Patrignani:2016xqp} to find the value quoted in Eq.~(\ref{fitco}).
This exceeds the fit results of Refs.~\cite{Ma:2010vd,Gong:2012ug} by roughly a
factor of two but is compatible with the lower bound extracted in
Ref.~\cite{Bodwin:2015iua}.
Further experimental and theoretical efforts are required to solve this
discrepancy.

\acknowledgments

We would like to thank M. Butensch\" on for useful discussions.
This work was supported in part by the German Federal Ministry for Education
and Research BMBF through Grant No.\ 05H15GUCC1 and by the China Scholarship
Council CSC through Grant No.\ CSC-201404910576.

\hfill \break
\onecolumngrid
\appendix
\section{Master Integrals at $\mathcal{O}(\epsilon)$}

In the following, we will drop all imaginary parts, since only the real parts
contribute to the NLO corrections. 
We adopt the notation of Ref.~\cite{Ellis:2007qk},
\beqn
 I_1^{D}(m_1^2) &=& \frac{\mu^{4-D}}{i\pi^{\frac{D}{2}}r_\Gamma} \int d^Dl \frac{1}{l^2-m_1^2 + i \varepsilon},
\nonumber\\
I_2^{D}(p_1^2;m_1^2,m_2^2) &=& \frac{\mu^{4-D}}{i\pi^{\frac{D}{2}}r_\Gamma}
 \int d^Dl \frac{1}{\left( l^2-m_1^2+i\varepsilon \right) \left[ (l+p_1)^2 - m_2^2 + i\varepsilon \right]},
\eeqn
where $r_\Gamma=\Gamma^2(1-\epsilon)\Gamma(1+\epsilon)/\Gamma(1-2\epsilon)
= 1 - \epsilon \gamma_E + \epsilon^2(\gamma_E^2/2 - 
\pi^2/12) + \mathcal{O}(\epsilon^3)$. 
The tadpole integral is given by
\beqn
I^{D}_1(m^2)  
=m^2\left(\frac{\mu^2}{m^2}\right)^\e \left[\frac{1}{\epsilon }+1+\e\left(1+\frac{\pi^2}{6}\right)\right].
\eeqn
The bubble integral with two vanishing masses is given by
\beqn
I_2^{D}(s;0,0)=\left(\frac{\mu^2}{s}\right)^\e\left[\frac{1}{\epsilon }+2 + \epsilon  \left(4-\frac{\pi^2}{2}\right)\right].
\eeqn
The bubble integral with one vanishing mass is given by
\beqn
I^{D}_2(s;0,m^2)  
=\left(\frac{\mu^2}{m^2}\right)^\e \left[\frac{1}{\e} - d_1 +\e\left(\frac{d_2}{2}+\frac{\pi^2}{6}\right)\right],
\eeqn
where 
\beqn
d_1=\frac{s-m^2}{s}\ln\frac{|m^2-s|}{m^2} -2,
\eeqn
and 
\beqn
d_2=2\frac{s-m^2}{s}\left(\ln{\frac{m^2-s}{m^2}}\ln{\frac{m^2-s}{|s|}}
-2\ln{\frac{m^2-s}{m^2}} - \text{Li}_2\left(\frac{m^2-s}{m^2}\right)
+\frac{\pi^2}{6}\right)+8
\eeqn
for $s<m^2$ and
\beqn
d_2=\frac{s-m^2}{s}\left(\ln^2{\frac{s-m^2}{m^2}}+\ln^2{\frac{s-m^2}{s}}
-4\ln{\frac{s-m^2}{m^2}} + 2 \text{Li}_2\left(\frac{s-m^2}{s}\right)
-\frac{5\pi^2}{3}\right)+8
\eeqn
for $s>m^2$.
The bubble integral with two different masses is given by
\beqn
I^{D}_2(\frac{t}{4};m_c^2,m_b^2)=\left(\frac{\mu^2}{t/4}\right)^\e\left[\frac{1}{\epsilon }-c_1 + \epsilon  \left(\frac{c_2}{2}+\frac{\pi ^2}{6}\right)\right],
\eeqn
where
\begin{eqnarray}
c_1&=&-\frac{1}{t}\left[-a\ln{\frac{s+u-a}{8m_cm_b}}+2\left(m_b^2-m_c^2\right)
  \ln{\frac{m_c^2}{m_b^2}}+t\ \ln{\frac{t}{4m_cm_b}}\right]-2,
\nonumber\\
c_2 &=& \frac{2a}{t}\left[\ln{\frac{s+u+a}{8m_c^2}}\left(\ln{\frac{a-s_b-u_b}{2a}}+\ln{\frac{t}{4m_b^2}}
+2\right)+\frac{1}{2}\ln^2{\frac{s+u+a}{8m_c^2}}+ \text{Li}_2\left(\frac{s_b+u_b+a}{2a}\right)\right.\nn\\
&&{}-\left.\text{Li}_2\left(\frac{-s_c-u_c+a}{2a}\right)\right]+\frac{s_c+u_c+a}{t}\ln{\frac{m_c^2}{m_b^2}}\left(\ln{\frac{t}{4m_cm_b}}+2\right)+\left(\ln{\frac{t}{4m_b^2}}+2\right)^2+4.
\end{eqnarray}

\section{Soft Integrals }

To extract the IR divergences of the real corrections in the limit
$k_5\rightarrow 0$, we need to know the results of the following integrals:
\begin{eqnarray}
  \int_{\text{soft}} \frac{d \text{PS}_{k_5} }{(K\cdot k_5)^2}&=&\frac{C_\e}{32 \pi^2 m_c^2} \left(-\frac{1}{\e}-\frac{s_c+u_c}{a}\ln{\frac{s_c+u_c+a}{s_c+u_c-a}} - \ln{\frac{\mu^2}{4\delta_s^2 m_c^2}} \right),
\nonumber\\
\int_{\text{soft}} \frac{d \text{PS}_{k_5} }{(k_3\cdot k_5)(k_4\cdot k_5)}&=&\frac{C_\e}{4 \pi^2 t}\left(\frac{1}{\e^2}+\frac{1}{\e}\ln{\frac{\mu^2}{4\delta_s^2m_c^2}}+\frac{1}{2}\ln^2{\frac{\mu^2}{4\delta_s^2m_c^2}}-\frac{\pi^2}{4}\right),
\nonumber\\
\int_{\text{soft}} \frac{d \text{PS}_{k_5} }{(K\cdot k_5)(k_3\cdot k_5)}&=&\frac{C_\e}{8 \pi^2s_c}\left[\frac{1}{\e^2}+\frac{1}{\e}\ln{\frac{\mu^2 t}{s_c^2\delta_s^2}}+\ln^2{\frac{s_c+u_c-a}{2s_c}}+\frac{1}{2}\ln^2{\frac{\mu^2}{4\delta_s^2m_c^2}}
-\ln{\frac{s_c^2}{4m_c^2t}}\ln{\frac{\mu^2}{4\delta_s^2m_c^2}}
\right.  \nn\\
  &&{}-\left.\frac{1}{2}\ln^2{\frac{s_c+u_c+a}{s_c+u_c-a}}+ 2\ \text{Li}_2\left(-\frac{s-u+a}{s_c+u_c-a}\right) -2\  \text{Li}_2\left(\frac{s-u-a}{2s_1}\right)-\frac{\pi^2}{4}\right] ,
\nonumber\\
\int_{\text{soft}} \frac{d \text{PS}_{k_5} }{(K\cdot k_5)(k_4\cdot k_5)}&=&\frac{C_\e}{8 \pi^2 u_c}\left[\frac{1}{\e^2}+\frac{1}{\e}\ln{\frac{\mu^2 t}{u_c^2\delta_s^2}}
  +\ln^2{\frac{s_c+u_c-a}{2u_c}}+\frac{1}{2}\ln^2{\frac{\mu^2}{4\delta_s^2m_c^2}}
-\ln{\frac{u_c^2}{4m_c^2t}}\ln{\frac{\mu^2}{4\delta_s^2m_c^2}}
\right.  \nn\\
  &&{}-\left.\frac{1}{2}\ln^2{\frac{s_c+u_c+a}{s_c+u_c-a}}+ 2\ \text{Li}_2\left(-\frac{u-s+a}{s_c+u_c-a}\right) -2\  \text{Li}_2\left(\frac{u-s-a}{2u_c}\right)-\frac{\pi^2}{4}\right].
\end{eqnarray}
\twocolumngrid

\end{document}